\documentclass[10pt]{elsarticle}
\usepackage{amssymb,amsmath}

\usepackage{hyperref}
\usepackage{color}

\def\bs{\boldsymbol}

\def\T{\te{T}}
\def\D{\te{D}}
\def\De{\te{D_e}}
\def\Dp{\te{D_p}}

\def\W{\te{W}}
\def\We{\te{W_e}}
\def\Wp{\te{W_p}}

\def\AA{\te{\mathcal{A}}}

\def\CC{\te{\mathcal{C}}}

\def\div{\mathrm{div\,}}

\newcommand\pd[2]{\frac{\partial #1}{\partial #2}}

\newcommand{\ve}[1]{\boldsymbol{#1}}
\newcommand{\te}[1]{\boldsymbol{#1}}
\newcommand{\ot}{\otimes}
\newcommand{\vv}{\te{v}}
\def\gradv{\nabla \vv}

\def\sym{\mathrm{sym}}

\def\L{\te{L}}
\def\Le{\te{L_e}}
\def\Lp{\te{L_p}}
\def\S{\te{S}}

\def\R{\te{R}}
\def\Fe{\te{F_e}}
\def\Fp{\te{F_p}}
\def\F{\te{F}}

\def\m{\ve{m}^{(\alpha)}}
\def\s{\ve{s}^{(\alpha)}}

\def\rho{\varrho}

\def\12{\frac{1}{2}}

\newcommand\ut[1]{\stackrel{\triangledown}{#1}}

\newcommand\be[2]{\begin{equation}\label{#1}#2 \end{equation}}

\journal{International Journal of Plasticity}    
    
\begin{document}
\begin{frontmatter}

\title{A Gibbs-potential-based framework for ideal plasticity of~crystalline solids treated as~a~material flow through an~adjustable crystal lattice space and its application to~three-dimensional micropillar compression}

\author[jk,jm]{Jan Kratochv\'\i l}
\author[jm]{Josef M\'alek}
\author[pm1,pm2]{Piotr Minakowski\corref{pmc}}
\cortext[pmc]{Corresponding author}
\ead{piotr.minakowski@iwr.uni-heidelberg.de}

\address[jk]{Czech Technical University, Faculty of Civil Engineering,  Th\'{a}kurova 7, 166~29 Prague 6, Czech Republic}
\address[jm]{Charles University in Prague, Faculty of Mathematics and Physics, Mathematical Institute, Sokolovsk\'{a} 83, 186~75 Prague 8, Czech Republic}
\address[pm1]{
Heidelberg University, Interdisciplinary Center for Scientific Computing, Im Neuenheimer Feld 205, 69120 Heidelberg, Germany}
\address[pm2]{University of Warsaw, Institute of Applied Mathematics and Mechanics, Banacha 2, 02-097 Warsaw, Poland}

\begin{abstract}
We propose an Eulerian thermodynamically compatible model for ideal plasticity of crystalline solids treated as a material flow through an adjustable crystal lattice space. The model is based on the additive splitting of the velocity gradient into the crystal lattice part and the plastic part. The approach extends a Gibbs-potential-based formulation developed in \cite{RajSr:11} for obtaining the response functions for elasto-visco-plastic crystals. The framework makes constitutive assumptions for two scalar functions: the Gibbs potential and the rate of dissipation. The constitutive equations relating the stress to kinematical quantities is then determined using the condition that the rate of dissipation is maximal providing that the relevant constraints are met. The proposed model is applied to three-dimensional micropillar compression, and its features, both on the level of modelling and computer simulations, are discussed and compared to relevant studies.
\end{abstract}


 \end{frontmatter}

\section{Introduction}

Severe plastic deformation experiments \cite{ValLang:06,ZhiLang:08} reveal
that a crystalline material at yield can be seen as  an anisotropic,
highly viscous fluid. A structural adjustment of the crystal lattice
space\footnote{The term ``space'' is used deliberately, as the crystal
lattice is understood as a space of preferred positions in a crystalline
phase considered regardless of particles staying or flowing through them.}
to the material flow is seen as  a deformation substructure. The flow
space is restricted to preferred crystallographic planes and the directions
causing anisotropy. Let us note that traditionally finite crystal
plasticity is based on a Lagrangian convected coordinate representation.
The presented form of crystal plasticity provides a possibility to treat
the flow-adjustment boundary-value problem alternatively as a fluid flow
in the Eulerian representation \cite{McGMcD:06}.

The traditional multiplicative split of the deformation gradient $\bs{F}$ into an elastic part $\Fe$ and a plastic part $\Fp$ in the form $\F=\Fe\Fp$ does not have a sound physical meaning. The sequential split means that the plastic action $\Fp$ occurs first, followed by the elastic action $\Fe$. However, that is not the way plasticity actually evolves; rather, elastic and plastic actions occur simultaneously. The formula $\F=\Fe\Fp$ is good for visualization purpose only, see  \cite{SrinivasaB2009}.

The additive splitting of the velocity gradient $\te{L}=\nabla \bs{v}$, where $\bs{v}$ is the velocity field,  into the elastic part $\Le$ and plastic part $\Lp$, i.e.,  
\begin{equation}\label{rule3}
\te{L} = \Le + \Lp\,.
\end{equation}
is far more natural and corresponds to the concept of plastic flow. The deformation gradient $\F$  can be found simply by solving the differential equation $\dot{\F}=\L\F$, where the superposed dot denotes material time-differentiation. In crystal plasticity, the parts $\Le$ and $\Lp$ are specified {in the following way: the elastic part $\Le$ is identified with the evolution of the lattice space vectors, while $\Lp$ is determined constitutively;
the Gibbs approach provides a suitable framework to achieve this, as will be shown in the next section.}

For modelling of crystal plasticity as a fluid  we adopt the Gibbs-potential-based framework proposed by Rajagopal and Shrinivasa \cite{RajSr:11}. They demonstrated the use of the Gibbs-potential-based formulation as a suitable tool for developing a thermodynamically consistent model for a wide class of elastic materials and viscoelastic fluids that are characterized by an implicit constitutive equation of the form
\begin{equation}\label{eq_ax_1}
\bs{f}(\bs{T}, \dot{\bs{T}},\bs{L}) = \te{0},
\end{equation}
where $\bs{T}$ is the Cauchy stress. 

For the class of models characterized through \eqref{eq_ax_1}, Rajagopal and Srinivasa \cite{RajSr:11} developed a thermodynamical framework that stems from the assumptions that the Gibbs potential is a function of the stress and the rate of dissipation is non-negative. In fact, the latter is strengthened by requiring that the rate of dissipation is maximum possible. The approach thus leads to models that satisfy the second law of thermodynamics, automatically.
%by starting with a stress dependent Gibbs potential, using the energy-dissipation equation and employing the requirement that the rate of mechanical dissipation to be non-negative. They strengthened this requirement by demanding that the dissipation rate is maximum possible. The approach accounts for a framework for generating models that satisfy the second law of thermodynamics, automatically. 
This thermodynamical framework, that is fully Eulerian (using the current state as a reference state), yields a constitutive equation of the form \eqref{eq_ax_1} from specifying the constitutive equations for two scalar functions: the Gibbs potential and the rate of dissipation. The importance of choosing the Gibbs potential from the set of thermodynamical potentials (including further the Helmholtz free energy, the internal energy and the enthalpy) stems from the fact that it allows one to use the stress as a primitive quantity instead of a kinematical quantity, as a measure of strain. As pointed out by Rajagopal and Srinivasa \cite{RajSr:11}, from a causal point of view, the traction (stress) is the cause, and the kinematics (motion or deformation) being the effect, which motivates to use of the stress as the primitive quantity. Requiring that the Gibbs potential is a function of a suitably chosen rotated stress tensor, Rajagopal and Srinivasa \cite{RajSr:11} extend their framework to model the response of anisotropic media. They  obtain rate type fluid models for both anisotropic elastic materials and for anisotropic viscoelastic materials of Maxwell type. 

The  objective of this paper is to refine and extend the Gibbs framework in order to  derive models that are capable of describing ideally plastic deformations of crystalline solids considered as a flow of material through an adjustable crystal lattice and that are furthermore compatible with the basic principles of continuum thermodynamics. The reason for focusing our attention to ideal plasticity, i.e. for ignoring work hardening or softening effects, is that an extension of the proposed model should not modify essential features of the Gibbs approach. To incorporate work hardening or softening would mean to enrich the model by suitable internal (history) variables governed by evolution equations (for a review of the hardening problem see e.g. \cite{SrinivasaB2009}). 

An Eulerian approach to dynamic crystal plasticity has been recently proposed, analysed and tested by Cazacu and Ionescu \cite{CazIon:09,Cazacu2010,Cazacu2010a}. Their goal was to develop an Eulerian rate-dependent model that is suitable for high strain rates, large strains and rotations of the incompressible material and the crystal lattice. Particular attention is devoted to the description of the kinematics of the crystal lattice in the Eulerian coordinates. The viscoplastic constitutive law and the equations for the evolution of the crystal lattice are expressed in terms of quantities attached to the current configuration.
In applications involving large deformations and high strain rates the elastic component of strain is small with respect to the inelastic one, and it can be therefore neglected. Using this argument a rigid-viscoplastic approach has been adopted. Modelling has been focused on in-plane deformation and on the role of inertia in the material response.

The novelty of the approach described herein is in ensuring that the considered models are compatible with the second law of thermodynamics, thanks to the fact that their derivation is based on the specification of constitutive assumptions for the Gibbs potential and the rate of dissipation, and by an application of the maximal rate of entropy production principle. In contrast with the common restriction to incompressible rigid body motions, the proposed model covers the case of fully elastic response. We also refrain from making the assumption that the material is incompressible since in processes such as high pressure torsion incompressibility seems to be a restrictive approximation. Compressibility, on the other hand, may relax non-physically high stresses in a 2-turn equal channel angular extrusion, cf.~\cite{Minak2turn}. 

We further illustrate the capabilities and the efficiency of the proposed model by performing numerical simulations for solving a three-dimensional micropillar compression problem. A material with face-centred cubic symmetry (FCC material) with 12 slip systems is considered. A numerical method is briefly outlined; more details and further numerical results confirming the efficiency of the numerical method will be given in a forthcoming paper.

The organization of the paper is as follows. In Subsection \ref{sec:gib}, we first recall the main ideas of the Gibbs-potential-based approach.  Then, in Subsection \ref{sec:CP}, we recall basic concepts of crystal plasticity, specify the forms of both the Gibbs potential and the rate of dissipation, and show that, together with balance equations, these constitutive equations for two scalar quantitities suffice for the derivation of the proposed elasto-visco-plastic  model of crystal plasticity. The derived model is then compared with the Eulerian approach of Cazacu and Ionescu \cite{Cazacu2010,Cazacu2010a} in  Section~\ref{sec:CI}. Section \ref{sec:sim} briefly introduces a numerical method and provides the results of numerical simulations for a micropillar compression ansatz. We conclude with a short overview of our main results in Section~\ref{sec:con}.

\section{A flow model of crystal plasticity}\label{sec:model}
\subsection{The Gibbs-potential-based formulation  by Rajagopal and Srinivasa \cite{RajSr:11}}\label{sec:gib}

In this section we summarise the result achieved in \cite{RajSr:11}. These results will be essential for our further consideration. 

We consider a body that occupies a configuration $\Omega_{t}$ at the current instant $t$. The current position of any particle is denoted by $\bs{x}$, its velocity by $\bs{v}$. Moreover, we introduce the symmetric and antisymmetric parts of the velocity gradient $\L = \gradv$ through
\begin{equation}
\!\!\!\!\!\! \D:= (\bs{L} + \bs{L}^{T})/2 \qquad \textrm{ and }\quad \bs{W}:= (\bs{L} - \bs{L}^{T})/2. \label{10a}
\end{equation}

The mass density of the material is denoted by $\varrho$ and the Cauchy stress by $\bs{T}$. The governing balance equations for mass, momentum and angular momentum (in the absence of external forces) are given in their local forms, as follows:
\begin{subequations}
\begin{eqnarray}
\dot \varrho +\varrho\, \text{div}\bs{v} &=& 0,\label{mass} \\
\varrho\, \dot{\bs{v}}  &=&  \text{div}\,\bs{T}, \qquad \bs{T}=\bs{T}^T\,, \label{balance}
\end{eqnarray}
\end{subequations}
where the superscript notation $\bs{A}^{T}$ means the transposed tensor to any tensor $\bs{A}$ and the material time derivative of a scalar function $z$ is given by $\dot{z}= z_{,t} + \nabla z \cdot \bs{v}$ (for a vector and tensor-valued function, the same relation is applied to each component). Furthermore, we will state the balance of energy (in the absence of the body heat supply) in the form
\begin{equation}
\varrho\, \dot{\epsilon} = \bs{T}:\D - \text{div}\,\bs{q}, \label{energy_balance}
\end{equation}
where $\epsilon$ is the specific internal energy, $\bs{q}$ the heat flux vector and $\D$ the symmetric part of the velocity gradient introduced in \eqref{10a}. Finally, we express the second law of thermodynamics through
\begin{equation}
\varrho\,\zeta:= \varrho\, \dot{\eta} + \text{div} \left(\frac{\bs{q}}{\theta}\right) \quad \textrm{ and } \quad \zeta\ge 0, \label{entropy_balance}
\end{equation}
where $\eta$ is the specific entropy, $\theta$ the temperature and $\zeta$ the specific rate of entropy production; here we tacitly assume that the entropy flux is of the form $\bs{q}/\theta$. 
Introducing the (specific) Helmholtz free energy $\psi$, the Kirchhoff stress tensor $\S$ and the (specific) rate of dissipation  $\xi$ through 
$$\psi := \epsilon - \theta\, \eta, \qquad  \S: =\bs{T}/\varrho \quad\textrm{ and } \quad \xi := \theta\,\zeta,$$
and using \eqref{energy_balance} and \eqref{entropy_balance}, we arrive at the equation for the rate of dissipation
\begin{equation}
\xi = \S:\D - \dot{\psi} - \eta \,\dot{\theta} - \bs{q}\cdot \frac{\nabla\theta}{\varrho\theta} \quad \textrm{ and } \quad \xi\ge 0. \label{entropy_balance_2}
\end{equation}
 
The starting point of the Gibbs-potential-based framework as developed in \cite{RajSr:11} is the assumption that the (specific\footnote{We suppress the use of the word \emph{specific} for relevant quantities in what follows although the quantities are taken per unit mass.}) Gibbs potential, denoted by $G$, is a function of the temperature $\theta$ and the Kirchhoff stress $\S$, i.e.,
\begin{equation}
G = {G}(\theta, \S) \qquad \textrm{ or } \qquad G(t,x) = {G}(\theta(t,x), \S(t,x)). \label{Gib_1}
\end{equation}
Stipulating further the Helmoltz free energy, the internal energy and the entropy, as functions of $\theta$ and $\S$, through 
\begin{equation}
\begin{split}
\psi(\theta, \S) &= G(\theta, \S) - \frac{\partial G (\theta,\S)}{\partial \S} : \S, \\
\epsilon(\theta, \S) &= G(\theta, \S) - \frac{\partial G(\theta, \S)}{\partial\S}:\S - \frac{\partial G(\theta, \S)}{\partial\theta}\theta, \\
\eta (\theta, \S) &= - \frac{\partial G(\theta, \S)}{\partial\theta}\theta,
\end{split}\label{int_energy}
\end{equation}
and inserting the first and third of these relations into \eqref{entropy_balance_2}, we obtain 
\begin{equation}
\xi = \S:\left\{\D + \frac{\partial^2 G}{\partial \S^2}\dot{\S} + \frac{\partial^2 G}{\partial \theta^2}\dot{\theta} \right\} - \bs{q}\cdot \frac{\nabla\theta}{\varrho\theta}\quad \textrm{ and } \quad \xi\ge 0. \label{entropy_balance_3}
\end{equation}

In what follows, we restrict ourselves to isothermal processes. Then, the equation \eqref{entropy_balance_3} reduces to  
\begin{equation}
\xi = \S:\D + \S:\frac{\partial^{2} G}{\partial\S^{2}}\dot{\S} \quad \textrm{ and } \quad \xi\ge 0. \label{dissipation}
\end{equation}

We have thus arrived at a representation of thermodynamics associated with the specification of the Gibbs potential (as given in \eqref{Gib_1}). There is however a problem: while $\D$ and $\S$ are both objective tensors, 
$\dot{\S}$ and consequently $({\partial^{2} G})/{(\partial\S^{2})}\, \dot{\S}$ are not objective tensors.

To overcome this difficulty and, moreover, to open the possibility to include anisotropic responses, Rajagopal and Srinivasa (see \cite{RajSr:11}) propose to consider, instead of \eqref{Gib_1}, the Gibbs potential depending on a rotated stress $\overline{\S}=\R^{T}\S\R$, i.e.,
\begin{equation}
G = G(\overline{\S}) = G(\R^{T}\S\R)\,, \label{anisotropicG}
\end{equation}
where $\R$ is any rotation tensor that is objective (which means that $\R$, related to a motion $\bs{\chi}$, and $\R^{*}$ related to the motion $\bs{\chi}^{*}$, satisfy $\R^{*} = \bs{Q}\R$ whenever $\bs{\chi}$ and $\bs{\chi}^*$ differ by a rigid body rotation $\bs{Q}$, all quantities being functions of position and time). 

Thus, assuming \eqref{anisotropicG} (instead of \eqref{Gib_1}) and following the same scheme to the one that yields \eqref{dissipation} from \eqref{Gib_1}, 
we get
\begin{equation}
\xi =  \S:\D + \overline{\S}:\frac{\partial^{2} G}{\partial\overline{\S}^{2}} \dot{\overline{\S}} = 
\S:\left\{\D - \R\overline{\AA} \, \dot{\overline{\S}}\R^{T}\right\} \quad \textrm{ and } \quad \xi\ge 0\,, \label{bardiss_rate}
\end{equation}
where 
\be{linearGibbs}{G(\overline{\S}) = -\12 \overline{\S}: \overline{\AA} \overline{\S} \text{ and } \overline{\AA} = - (\partial^{2} G)/(\partial\overline{\S}^{2}).}

Substituting $\overline{\S} = \R^{T}\S\R$ into the expression on the right-hand side of (\ref{bardiss_rate}) and using the fact that $\R^{T}\R = \bs{I}$ we obtain 
\begin{equation*}
\begin{split}
\S: \left\{\D - \R\overline{\AA} \, \dot{\overline{\S}}\R^{T}\right\} &= \S:\left\{\D - \R\overline{\AA} \, \R^{T} \left(\dot{\S} + \R \dot{\R}^{T} \S + \S \dot{\R} \R^{T} \right) \R \R^{T}\right\}\\
&= \S:\left\{\D - \AA (\dot{\S} + \S \dot{\R} \R^{T} + (\dot{\R} \R^{T})^{T} \S) \right\}\\
&= \S:\left\{\D - \AA (\dot{\S} + \S \bs{\Omega} - \bs{\Omega} \S) \right\},
\end{split}
\end{equation*}
where $\AA$ is a fourth-order tensor such that $\mathcal{A}_{ijkl} = \R_{iA}\R_{jB}\R_{kC}\R_{lD} \overline{\mathcal{A}}_{ABCD} $ and $\bs{\Omega}=\dot{\R}\R^{T}$. Note that $\R\R^{T} = \bs{I}$ implies that $\dot{\R}\R^{T}=-\R\dot{\R}^{T} = - (\dot{\R}\R^{T})^{T}$ and $\bs{\Omega}$ is antisymmetric. Next, introducing the notation 
\begin{equation}\label{objectiverate1stdef}
\ut{\S}:=\dot{\S}+\S\bs{\Omega}-\bs{\Omega}\S\,,
\end{equation}
it is not difficult to observe that $\ut{\S}$ is an objective time derivative. With this notation, \eqref{bardiss_rate} takes the form 
\begin{equation}
\xi =  \S:(\D - \AA\ut{\S}) \quad \textrm{ and } \quad \xi\ge 0\,, \label{objectiverate}
\end{equation}
where $\S$, $\D$ and $\AA\ut{\S}$ are objective and symmetric second-order tensors. In addition, this step provides a simple and natural way for introducing, through $\AA$, an anisotropic response as there is no restriction imposed by frame indifference on the dependence of $G$ on $\overline{\S}$. 
%minak removed to discussed in more details in the cp section
%By choosing different rotated stress tensors, one ends up with different objective time derivatives. For example, by choosing $\R$ such that $\dot{\R}\R^{T} = \bs{W}$ (or equivalently $\dot{\R} = \bs{W}\R$), we obtain the Zaremba-Jaumann derivative of $\S$, i.e. $\dot{\S} + \S\bs{W} - \bs{W}\S$. 

Rajagopal and Srinivasa \cite{RajSr:11} use \eqref{objectiverate} to make, among others, the following  observation: the dissipation rate $\xi$ vanishes for arbitrary $\S$ if 
\begin{equation}
\D = \AA\ut{\S} \quad \textrm{ i.e. } \quad \D = \AA\{\dot{\S}+\S\bs{\Omega}-\bs{\Omega}\S \}\label{eq_ax_2}.
\end{equation}
This constitutive equation, which is of the form \eqref{eq_ax_1}, characterizes elastic (nondissipative) materials \footnote{ As is well known such elastic response cannot be achieved via the Helmholtz free energy approach, see \cite{RajSr:11} for further details and references.}. In the next Section, the Gibbs potential (\ref{anisotropicG}) and the constitutive assumption (\ref{eq_ax_2}) modified for elastic properties of the crystal lattice space will be used for modeling of crystal plasticity.

\subsection{Crystal plasticity}\label{sec:CP}

A formulation of crystal plasticity in the framework of the Gibbs-potential approach is derived in two steps. First, we recall the standard description of a crystal lattice space as well as slip systems and identify the lattice velocity gradient ${\bs L}_{e}$ with the evolution of the lattice space. As a second step we specify the constitutive equation for the rate of dissipation and by applying the assumption that the response of the material is such that it maximizes this dissipation function (and other relevant constraints) we obtain the constitutive equation for $\S$. 

A crystal lattice space can be characterized by three non-planar vectors $\bs{a}_{1}$, $\bs{a}_{2}$ and $\bs{a}_{3}$, which form a basis in $\mathbb{R}^{3}$ and span the so-called Bravais lattice (as an example, a face-centred cubic lattice is sketched in Fig. 1a). The reciprocal dual lattice vectors are defined by 
\begin{equation} \label{reciprocal}
\ve{a^1} = \frac{\ve{a_2}\times \ve{a_3} }{\ve{a_1}\cdot (\ve{a_2}\times \ve{a_3}) },
\end{equation}
with $\bs{a}^{2}$ and $\bs{a}^{3}$ defined by cyclic permutation; note that $\bs{a}^{i}\cdot\bs{a}_{j}=\delta_{ij}$. 

\begin{figure}[h!]
\begin{center}
\includegraphics[scale=1.]{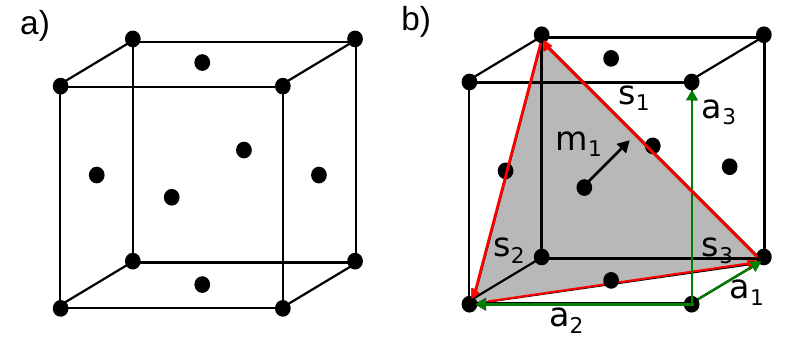}
 \end{center}
\caption{Face-centred  cubic (FCC) crystal structure is sketched in the left image a), while an example of the slip system, i.e. the plane along which the crystal slides, is shown in grey in the right image b). }
\label{fig:cp_basis}
\end{figure}

In crystalline solids the material flow in a plastic regime is carried by a motion of dislocations. Crystal plasticity theory is based on the observation that the motion of dislocations takes place in preferred crystallographic directions (so-called slip directions) on preferred crystallographic planes (so-called slip planes); in our consideration climb of dislocations is excluded. The unit vector in the slip direction $\bs{s}^{(\alpha)}$ and the unit normal to the slip plane $\bs{m}^{(\alpha)}$ form the $\alpha$-slip system. The set of $N$ potentially active slip systems is one of the basic microscopic characteristics of crystal plasticity, hence $\alpha= 1,\dots,N$. The vectors $\bs{s}^{(\alpha)}$ and $\bs{m}^{(\alpha)}$ of the slip system $\alpha$ can be expressed through the lattice vectors $\bs{a}_{1}$, $\bs{a}_{2}$ and $\bs{a}_{3}$ as
\begin{equation} \label{slipsystem}
\bs{s}^{(\alpha)}=\sum_{i=1}^3s^{\alpha}_{i}\,\ve{a_i}\,, \quad \bs{m}^{(\alpha)}=\sum_{i=1}^3m^{\alpha}_{i}\,\bs{a}^{i},
\end{equation}  
where $s^{\alpha}_{i}$ and $m^{\alpha}_{i}$ are constant coefficients, called Miller indexes. For example, for $s^{1}_{1}=1$, $s^{1}_{2}=1$ and $s^{1}_{3}=0$, and $m^{1}_{1}=1$, $m^{1}_{2}=1$ and $m^{1}_{3}=1$ the slip direction vector $\bs{s}^{1}=\bs{a}_{1} +\bs{a}_{2}$ and the slip plane normal $\bs{m}^{(1)}=\bs{a}_{1} +\bs{a}_{2} +\bs{a}_{3}$, expressed in a standard way through the Miller indexes, read [110] and (111), respectively; [110](111) is a typical slip system of face-centred cubic crystals as seen in Fig. \ref{fig:cp_basis}b). (Negative values are denoted by bars above the (positive) numbers.)

In a deformation process the crystal lattice space adjusts to the stress and the material flow.  In the proposed model, the crystal lattice is treated as a solid described by the lattice deformation gradient $\Fe$ from the natural (lattice reference) configuration \footnote{The concept of the natural configuration has been developed and applied in a number of areas beyond the area of plasticity by Rajagopal and his co-workers, see for example \cite{Rajagopal1998a, Rajagopal1998b, Rajagopal2004}. In our setting, the natural configuration is the lattice reference configuration and thus  
\begin{equation} \label{Fe}
\ve{a_i}(\bs{x},t)=\Fe(\bs{x},t)\,\ve{a_i}(\bs{x},0)\,,
\end{equation} 
where $\ve{a_i}(\bs{x},0)$ are the lattice vectors in the natural (lattice reference) configuration.

Since the lattice velocity gradient $\Le$ is supposed to be linked with the rate of $\Fe$ through the relation $\Le = \dot\Fe\Fe^{-1}$ (that is analogous to the kinematical relationship between $\L$ and $\dot\F$), the rate form of (\ref{Fe}) reads
\begin{equation} \label{rateFe}
{\dot{\bs{a}}_{i}}(\bs{x},t)=\Le\,\ve{a_i}(\bs{x},t)\,.
\end{equation}}  to the current configuration. According to Srinivasa and Srinivasan \cite[Section 13.4, p. 467]{SrinivasaB2009} this evolution equation for the lattice vectors $\ve{a_i}$ is characterised in the following way
\begin{equation} \label{latticerate}
\Le = \sum_{i=1}^3 {\dot{\ve{a}}_{\ve{i}}}\otimes \ve{a^i} \iff \dot{\ve{a}}_{\ve{j}} = \Le \ve{a_j}. 
\end{equation}
Indeed, to see that \eqref{latticerate} holds, we first multiply $\sum_{i=1}^3 \dot{\ve{a}}_{\ve{i}}\otimes \ve{a^i}$ by $\ve{a_j}$  ($ \ve{a^i} \cdot \ve{a_j} = \delta_{ij}$) and observe that $\Le =  \sum_{i=1}^3 \dot{\ve{a}}_{\ve{i}} \otimes \ve{a^i} \Rightarrow \dot{\ve{a}}_{\ve{j}} = \Le \ve{a_j}.$ On the other hand, the opposite implication, i.e.  $\Le =  \sum_{i=1}^3 \dot{\ve{a}}_{\ve{i}}\otimes \ve{a^i} \Leftarrow \dot{ \ve{a_j}} = \Le \ve{a_j}$, follows from the definition of the reciprocal basis \eqref{reciprocal}.
Note that the lattice vectors remain orthogonal, since they are transformed by the same lattice deformation gradient~$\Fe$\footnote{
We decompose the lattice deformation gradient $\Fe$ into the elastic stretch $\te{U_{e}}$ and the rotation $\te{R_e}$, $\Fe = \te{R_e}\te{U_{e}}$. The angle is preserved by the rotation $\te{R_e}$, and the loss of orthonormality is caused by the elastic stretching.}.

%{ \color{blue} Piotr question: Is this clear or we should transfer argument to the rate form?}

Our basic kinematical assumption is represented by the  additive splitting of the velocity gradient in the form (\ref{rule3}), i.e. 
\begin{equation}\label{rule4}
\te{L} = \Le + \Lp\,,
\end{equation}
where $\Le$ is linked with the $\ve{a}_{\ve{i}}$ via the equations \eqref{latticerate} and  the form of $\Lp$ will be specified by maximization of the rate of dissipation based on the identification of the constitutive equation for the rate of dissipation. 

Following the decomposition \eqref{rule4}, we set 
\begin{align}
\De&:= (\Le + \Le^T)/2 \quad\textrm{ and } \quad \We:= (\Le - \Le^T)/2, \label{10b}\\
\Dp&:= (\Lp + \Lp^T)/2 \quad \textrm{ and } \quad \Wp:= (\Lp - \Lp^T)/2,\label{10c}
\end{align}
and from (\ref{10a}) we have that
\begin{equation}
\D= \De + \Dp \quad\textrm{ and } \quad \W= \We + \Wp. \label{10d}
\end{equation}

Considering \eqref{anisotropicG} and \eqref{objectiverate} with the constitutive equation (\ref{eq_ax_2}) modified to the crystal lattice space $\De = \AA \, \ut{\S}$ we get from  \eqref{latticerate} that 
\begin{equation} \label{latticerate1}
\De = \AA\ut{\S} = \textrm{sym}\left( \sum_{i=1}^3 \ve{\dot a_i}\otimes \ve{a^i} \right) \quad \textrm{ and } \quad \Le = \sum_{i=1}^3 \ve{\dot a_i}\otimes \ve{a^i}.
\end{equation}

Next, we employ the procedure of maximization of the rate of dissipation $\xi$ to get $\D_{p}$; this, together with the second equation in \eqref{latticerate}, also allows us to determine uniquely the rate of material flow $\Lp$ for a crystalline solid deformed by slip. To do so, we need to specify the constitutive equation for the rate of dissipation $\xi$. The driving force of slip is the resolved shear stress $\tau^{(\alpha)}$ controlled by the critical value $\tau^{(\alpha)}_{c}$ (\cite{SrinivasaB2009}), 
\begin{equation} \label{tau}
\tau^{(\alpha)} = \bs{s}^{(\alpha)}\cdot \S\,\bs{m}^{(\alpha)}\,.	
\end{equation}
For the sake of the following procedure the critical values $\tau^{(\alpha)}_{c}$ are assumed to be a given positive material parameter in agreement with our restriction to ideal plasticity. If $\tau^{(\alpha)}<\tau^{(\alpha)}_{c}$ the slip rate and the corresponding dissipation rate of the $\alpha$-slip system is very small. On the other hand, if $\tau^{(\alpha)}>\tau^{(\alpha)}_{c}$ the dissipation rate is high. One of the suitable possibilities is to assume that the dissipation rate is a sum of contributions of the slip systems proportional to a power of the quotients $\tau^{(\alpha)}/\tau^{(\alpha)}_{c}$, i.e.,
\begin{equation} \label{xi}
	\hat{\xi}(\S)  = \sum_{\alpha=1}^{N} c_{0} \left(\frac{|\tau^{(\alpha)}|}{\tau^{(\alpha)}_{c}}\right)^{\beta+1} = \sum_{\alpha=1}^{N} c_{0} \left(\frac{|\bs{s}^{(\alpha)}\cdot \S\,\bs{m}^{(\alpha)}|}{\tau^{(\alpha)}_{c}}\right)^{\beta+1}\,,
\end{equation}
where $c_{0}>0$ and $\beta>0$ are material constants. 
 
Introducing the notation
\begin{equation} \label{alpha slip}
\nu^{(\alpha)} : = c_{0}\,\frac{\text{sgn}(\tau^{(\alpha)})}{\tau^{(\alpha)}_{c}} \left(\frac{|\tau^{(\alpha)}|}{\tau^{(\alpha)}_{c}}\right)^{\beta}\,,	
\end{equation}
we observe that  
\begin{equation}\label{eq-aux-3}
\pd{\hat\xi(\S)}{\S} = (\beta+1) \sum_{\alpha=1}^{N} \nu^{(\alpha)} \text{sym}(\bs{s}^{(\alpha)} \otimes \bs{m}^{(\alpha)}) 
\end{equation}

and consequently 
\begin{equation}\label{eq-aux-2} 
 \pd{\hat\xi(\S)}{\S}:\S = (\beta+1)\hat{\xi}(\S). 
\end{equation}

In order to determine the constitutive equation for the evolution of $\S$, we maximize $\hat\xi$ given in \eqref{xi} with respect to $\S$ under the constraint \eqref{objectiverate} that can be rewritten in a more compact form, namely 
\be{constnew}{\xi = \S:\{\D - \De\} = \S:\Dp \quad \textrm{ and } \xi \ge 0.} Hence, setting 
$$L(\S) := \hat{\xi}(\S) + \ell (\hat{\xi}(\S)-\S:\Dp)\,,$$ 
the constrained optimality condition reads
\begin{equation}\label{eq-aux-5}
0=\pd{L(\S)}{\S}=(1+\ell) \pd{\hat\xi(\S)}{\S} -\ell \Dp \quad \iff \quad \Dp = \frac{1+\ell}{\ell}\pd{\hat\xi(\S)}{\S}\,.
\end{equation}
Multiplying the resulting equation scalarly by $\S$, and using \eqref{constnew} and \eqref{eq-aux-2}, we conclude that 
$$
  \frac{1+\ell}{\ell} = \frac{1}{\beta+1}\,.
$$
Consequently, the second condition in \eqref{eq-aux-5} together with \eqref{latticerate1} and \eqref{eq-aux-3} imply that
\begin{equation}\label{eq-aux-6}
\D - \AA\ut{\S} = \Dp = \sum_{\alpha=1}^{N} \nu^{(\alpha)} \text{sym}(\bs{s}^{(\alpha)} \otimes \bs{m}^{(\alpha)})\,.
\end{equation}
Note that $\nu^{(\alpha)}$ can be interpreted as a slip rate of the $\alpha$-slip system.
Thus, we have identified $\Dp$. Since \eqref{latticerate} and \eqref{10b} imply that $\W = \We +  \Wp$ uniquely determines $\Wp$, we stipulate that 
\be{traditional}{
	\Lp = \sum_{\alpha=1}^{N}\nu^{(\alpha)}(\bs{s}^{(\alpha)}\otimes\bs{m}^{(\alpha)})\,.
}
Note that this formula is consistent with the results obtained so far for the symmetric parts $\D$, $\D_e$ and $\D_p$ since we have (see \eqref{eq-aux-6}) 
\be{41new}{\D = \De + \D_p = \AA\ut{\S} + \sum_{\alpha=1}^{N} \nu^{(\alpha)} \text{sym}(\bs{s}^{(\alpha)} \otimes \bs{m}^{(\alpha)})\,.
}
In addition, the choice \eqref{traditional} and the decomposition rule (\ref{rule4}) lead~to 
\begin{equation} \label{basicrule}
	\nabla \bs{v} =\sum_{i=1}^3 \dot{\bs{a}}_{i}
\otimes  \bs{a}^{i} + \sum_{\alpha=1}^{N}\nu^{(\alpha)}(\bs{s}^{(\alpha)}\otimes\bs{m}^{(\alpha)}),
\end{equation}
which we view (and use) as the evolutionary equation for the vectors $\bs{a}_i$.

Recalling definition \eqref{objectiverate1stdef} and combining it with the obtained constitutive relation \eqref{eq-aux-6} we obtain

$$\D - \AA\{\dot{\S}+\S\bs{\Omega}-\bs{\Omega}\S \} = \sum_{\alpha=1}^{N} \nu^{(\alpha)} \text{sym}(\bs{s}^{(\alpha)} \otimes \bs{m}^{(\alpha)}),$$
where $\te{\Omega} = \dot\R\R^T$ for any rotation tensor $\R$ that is objective. By choosing different rotations $\R$, one ends up with different objective time derivatives, and, consequently, with different models, with the restriction for $\te{\Omega}$ being antisymmetric. For example, by choosing $\R$ as the rotational part $\te{R_e}$ of the polar decomposition of the lattice deformation gradient $\Fe = \te{R_e}\te{U_{e}}$ will yield the Green-McInnis-Naghdi stress rate. The second example, namely the Zaremba-Jaumann derivative, extensively used in the crystal plasticity literature, results from setting $\te{\Omega} = \dot\R\R^T = \We$. 

From now on we focus on the Zaremba-Jaumann rate
\begin{equation}
\ut{\S}=\dot{\S}+\S\We-\We\S.\label{10e}
\end{equation}

At this point we compare the constitutive relation \eqref{eq-aux-6} with relations that are typically used in crystal plasticity, see e.g. \cite{MinakMetz, Raabe2000}. In these studies, based on the Helmholtz potential approach, the resulting constitutive equation takes the form 
\be{helholtzrate}{\dot\S - \te{W_e}\S- \De\S + \S\te{W_e} - \S\De= \CC(\De),} 
where the rate derivative on the left-hand side is the Oldroyd derivative. To obtain the Zaremba-Jaumann rate, the terms $\De\S + \S\De$ are often neglected or transferred to the right-hand side of \eqref{helholtzrate} by defining the fourth-order tensor $\mathcal{\te{K}}(\De) = \CC(\De) + \De\S + \S\De$. In this sense the Gibbs and the Helmholtz potential based approaches are comparable. %The Oldroyd derivative can be recovered within the proposed framework by modifying assumptions on $\AA$, however this requires additional work. 
%{ \color{red}  J. Kratochvil comment: However, the perspective described in this work is more general. $->$ put it away or specify }
%{ \color{blue} Piotr answer: I thought that this is clear from previous paragraph that Helmholtz is a specific case of Gibbs and therefore Gibbs is more general.}

\subsection{The model  and its comparison to the Eulerian approach by Cazacu and Ionescu}\label{sec:CI}
In summary, the primary variables of the proposed crystal plasticity model based on the Gibbs potential are: the density $\varrho$, the velocity $\bs{v}$, the Kirchhoff stress $\S$, and the vectors $\bs{a}_{1}$, $\bs{a}_{2}$ and $\bs{a}_{3}$ being functions of the current position $\bs{x}$ and time $t$. These variables are governed by the system of equations (\ref{mass}), (\ref{balance}), \eqref{eq-aux-6} and (\ref{basicrule}), i.e.,

\begin{subequations}\label{finalsystem}
\begin{eqnarray}
\dot \varrho +\varrho\, \text{div}\,\bs{v} &=& 0,\label{finalsystema} \\
\varrho\, \dot{\bs{v}}  &=&  \text{div}\,(\varrho \S)\,, \label{finalsystemb}\\
\AA\ut{\S} &=& \sum_{i=1}^3 \text{sym}(\dot{\bs{a}}_{i}\otimes\bs{a}^{i})\,, \label{finalsystemc} \\
\sum_{i=1}^3 \dot{\bs{a}}_{i}
\otimes \bs{a}^{i} &=& \nabla \bs{v} - \sum_{\alpha=1}^{N}\nu^{(\alpha)}(\bs{s}^{(\alpha)}\otimes\bs{m}^{(\alpha)})\,. \label{finalsystemd}
\end{eqnarray}
\end{subequations}

The slip system vectors $\bs{s}^{(\alpha)}$ and $\bs{m}^{(\alpha)}$ in (\ref{finalsystemd}) are expressed in terms of the lattice vectors $\bs{a}_{1}$, $\bs{a}_{2}$ and $\bs{a}_{3}$ through  (\ref{reciprocal}) and (\ref{slipsystem}); the slip rates $\nu^{(\alpha)}$ are governed by the equations (\ref{alpha slip}), (\ref{tau}). 

We also multiply \eqref{finalsystemc} by $\CC := \AA^{-1}$ and obtain
\be{C}{\ut{\S} = \CC\sum_{i=1}^3 \text{sym}(\dot{\bs{a}}_{i}\otimes\bs{a}^{i}) = \CC\De .}
The elastic tensor $\CC$ is anisotropic; in what follows we assume that $\CC$ is of cubic symmetry characterized by three independent elements, $c_{11}$, $c_{12}$, and $c_{44}$, which correspond to $\mathcal{C}_{1111}$, $\mathcal{C}_{1122}$, and $\mathcal{C}_{2332}$ respectively, through the standard Voigt notation.

In comparison with (\ref{finalsystem}) the primary variables of the rigid-plastic model  proposed by Cazacu and Ionescu \cite{CazIon:09,Cazacu2010a,Cazacu2010} are: the velocity $\bs{v}$, which is divergence-free, the Cauchy stress $\bs{T}=\varrho\S$, and the slip system vectors $\bs{s}^{(\alpha)}$ and $\bs{m}^{(\alpha)}$, $\alpha=1,\ldots,N$ as functions of the current position $\bs{x}$ and time $t$. These variables are governed by the system of the equations:
\begin{subequations} \label{CI}
\begin{eqnarray}
 \text{div}\,\bs{v} &=& 0,\label{CIa} \\
\varrho\, \dot{\bs{v}}  &=&  \text{div}\,\bs{T}\,, \label{CIb}\\
\D &=& \sum_{\alpha=1}^N \nu^{(\alpha)} \text{sym}(\bs{s}^{(\alpha)}\otimes\bs{m}^{(\alpha)})\,, \label{CIc} \\
 \dot{\bs{s}}^{(\alpha)} &=& (\We - \sum_{\alpha=1}^N \nu^{(\alpha)} \text{skew}(\bs{s}^{(\alpha)}\otimes\bs{m}^{(\alpha)})) \bs{s}^{(\alpha)}\,, \label{CId}\\
\dot{\bs{m}}^{(\alpha)} &=& (\We - \sum_{\alpha=1}^N \nu^{(\alpha)} \text{skew}(\bs{s}^{(\alpha)}\otimes\bs{m}^{(\alpha)})) \bs{m}^{(\alpha)}\,. \label{CIe}
\end{eqnarray}
\end{subequations}
The expression in the brackets on the right-hand side of (\ref{CId}) and (\ref{CIe}) is the rate of the lattice rotation $\dot{\R}_{e}\R_{e}^{-1}$ (for a rigid lattice $\Fe=\R_{e}$); (\ref{CId}) and (\ref{CIe}) are the rigid plasticity versions of (\ref{finalsystemd}) of our model evaluated directly in terms of the vectors of the slip system $\bs{s}^{(\alpha)}$ and $\bs{m}^{(\alpha)}$, $\alpha= 1,\ldots,N$ instead of the lattice vectors $\ve{a_i}$, $i=1,2,3$ in our case. Since $N$ can be large (for e.g. for FCC crystal structure $N=12$), working with the lattice vectors $\ve{a_i}$ makes the system \eqref{finalsystem} advantageous with respect to the system \eqref{CI}. It reduces the number of unknowns in the system and thus plays an important role in the efficiency of three-dimensional numerical simulations.

Due to the assumed rigidity, the density $\varrho$ in (\ref{CIb}) is a material constant. The equation (\ref{CIc}) is an implicit constitutive equation for the stress $\T$ as could be seen if one used for the slip rates $\nu^{(\alpha)}$ the power law (\ref{alpha slip}) and for the resolved shear stress $\tau^{(\alpha)} = \bs{s}^{(\alpha)}\cdot\bs{T}\,\bs{m}^{(\alpha)}$. Note that in this case all $N$ slip systems are active, and the power law determines their relative activity. However, Cazacu and Ionescu preferred, mainly from the computational point of view,
to employ instead the power law (\ref{alpha slip}) the visco-plastic extension of the classical rigid-plastic Schmid law using the overstress approach  
\begin{equation} \label{overstress}
\nu^{(\alpha)} = \frac{1}{\eta^{(\alpha)}}[|\tau^{(\alpha)}|-\tau^{(\alpha)}_{c}]_{+}\text{sign} (\tau^{(\alpha)})\,,	
\end{equation}
where $\eta^{\alpha}$ is the viscosity; $[z]_{+} = (z+|z|)/2$ denotes the positive part of the real number $z$. Using this overstress approach, it is important to mention that only up to 5 resolved shear stresses $\tau^{(\alpha)}$ can be independent (there are 5 components of the stress deviator controlling slip). The slip rates $\nu^{(\alpha)}$ given by (\ref{overstress}) are also not independent as they have to satisfy the kinematic constrain (\ref{CIc}). Given the deformation rate $\D$, the slip rate $\nu^{(\alpha)}$ of the active slip systems can be determined by minimizing the internal power under the constrain (\ref{CIc}). The deviatoric part of the stress $\bs{T}$ corresponding to a given deformation rate $\D$ is obtained as the Lagrange multiplier of this minimization problem. 

One of the main differences is that in the rigid-plastic model (\ref{CI}) the elastic properties of the lattice are neglected as they are assumed to be small in comparison with the large plastic deformations. This assumption is reasonable in modelling e.g. metal production processes and generally it is suitable from a macroscopic point of view. If the interest is focused on a meso-scale, where the deformation is spontaneously heterogeneous, not only the local misorientations of the crystal lattice but also lattice strains connected with a stress redistribution may play a role. It is worth mentioning that classical crystal plasticity models of shear bands consider anisotropic elasticity as an important ingredient. Moreover, elasticity and compressibility have a stabilizing effect, helping the computational strategy as indicated in Section \ref{sec:sim}, see also \cite{Minak2turn}.

In some aspects the models (\ref{finalsystem}) and (\ref{CI}) are similar. Both use the rate decomposition rule (\ref{rule4}) in the current configuration as the basic kinematic assumption (the reference to decomposition rule (\ref{rule4}) in Cazacu and Ionescu \cite{Cazacu2010} is only indicative) and their Eulerian descriptions are conceptually equivalent (except the acceptance of elasticity and the difference in the power vs. overstress rules mentioned above). A substantial difference is in the formulations of the models. The rigid-plastic model (\ref{CI}) is based on the slip rate composition of the plastic velocity gradient (\ref{traditional}) and the resolved shear stress (\ref{tau}) as the primary assumptions. On the other hand, the proposed constitutive modelling based on the Gibbs-potential formulation is constructed from two scalar functions of the Kirchhoff stress $\S$: the Gibbs potential $G(\overline{\S})$ given by (\ref{linearGibbs}), $\overline{\S}= \R_{e}^{T}\S\R_{e}$, and the rate of dissipation $\xi(\S)$ given by (\ref{xi}). The form of the material flow gradient $\Lp$ given by (\ref{traditional}) is the consequence of the dissipation rate maximization. Moreover, the Gibbs-potential framework guarantees that the model is thermodynamically consistent. 

\section{Numerical example: three-dimensional micropillar compression}\label{sec:sim}

%New paragraph 
The algorithm presented in this section is based on the Eulerian framework described above and thus contributes to the Crystal Plasticity Finite Element Method (CPFEM). A recent overview by Roters et al. \cite{Roters2010, book:Roters2010} summarizes applications of the CPFEM. Most of the methods that have been developed so far adopt a Lagrangian description of the continuum problem. In the context of the present paper we specifically highlight the following references focused on the application of CPFEM to micropillar compression: \cite{Hurtado2012, Hurtado2013, Kuroda2013, Jung2015, Soler2012}. 

In \cite{Jung2015}, Jung et al. compare different primary slip plane inclination angles and use a hardening rule that accounts for anisotropic slip system interactions. Currently, it seems that the main interest in the field concentrates on finite element analysis of geometrically necessary dislocations, see e.g. \cite{Hurtado2012, Hurtado2013, Kuroda2013}. Since the primary purpose of this study is to computationally analyze the Eulerian approach based on the Gibbs-potential-based formulation, we focus on the case of ideal plasticity here and postpone the study of the hardening (and possibly non-local) effects to our further research. Note, however that in Minakowski et al. \cite{MinakMetz} the authors presented a crystal plasticity model including isotropic hardening and proposed an Arbitrary Lagrangian Eulerian (ALE) based numerical method  for a two-dimensional plastic flow of a single crystal compressed in a channel die.

To examine the potential of the derived model we consider three-dimensional micropillar compression. For comparison we use the same setup as Kuroda in \cite{Kuroda2013}, where the behaviour of single crystal micropillars subjected to compressive loading is described. A three-dimensional finite element method incorporating higher-order gradient crystal plasticity is analyzed. The author first simulated a compression test for a model without higher strain gradients, which we use for comparison. 

\begin{figure}[!ht]
\begin{center}
      \setlength{\unitlength}{.1\textwidth}
      \begin{picture}(10,5)
        \put(0.15,0){\includegraphics[width=\textwidth]{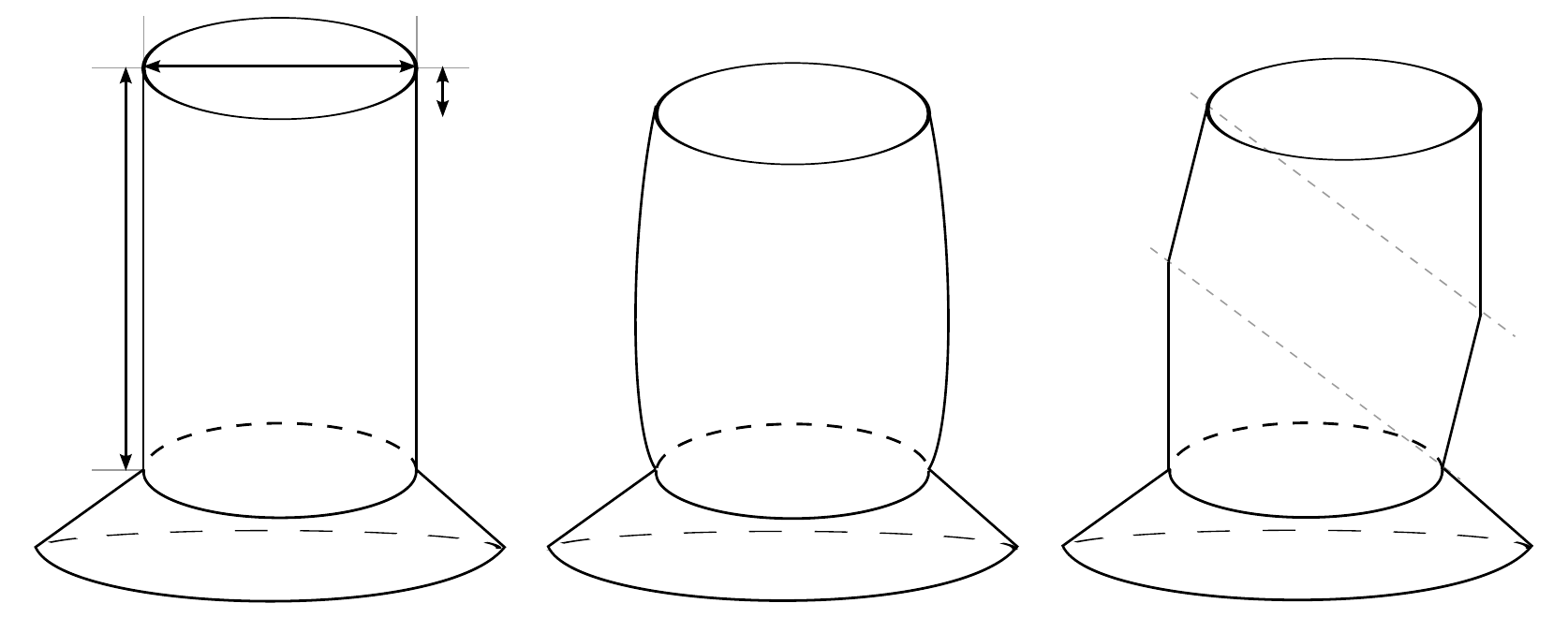}}
        \put(0.5,2.){$L_0$}
        \put(1.9,3.55){$D$}
        \put(3.1,3.25){$L$}
        \put(0,4.){$(a)$}
        \put(3.5,4){$(b)$}
        \put(7,4){$(c)$}
        \put(1.9,2.9){$\Gamma_1$}
        \put(1.9,0.8){$\Gamma_3$}
        \put(2.4,2.){$\Gamma_2$}
     \end{picture}
\end{center}
\caption{Scheme of micropillar compression (a),  compressed sample with constrained (b) and unconstrained (c) boundary conditions.}
\label{fig:3dcomp}
\end{figure}
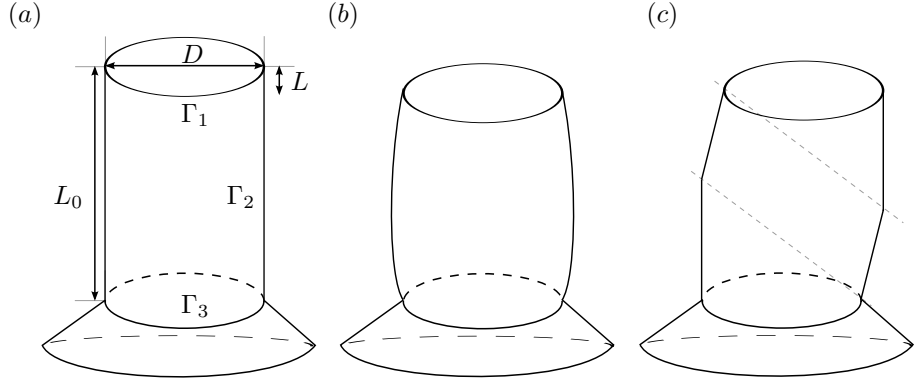

The compression testing methodology is shown schematically in Figure \ref{fig:3dcomp} (like Kuroda \cite{Kuroda2013} we consider samples with a conical base). The setting mimics the compression experiment commonly performed on macroscopic samples \cite{Uchic2004, Uchic2009, Shade2009}. The compressive stress is defined as the sum of the nodal forces in the $-x_{3}$ direction at the top surface divided by the initial cross-section area. The nominal compressive strain is the ratio of the displacement of the top surface in the $-x_{3}$ direction over the original height of the sample.

We use the same set of values of parameters of the compressed material as in \cite{Kuroda2013}, which corresponds to the experimental and numerical works   \cite{Uchic2007, Shade2009}. The ratio between the height $L_0$ and the diameter $D$ is selected to be $2.3$. In a crystal with cubic symmetry, such as face-centred cubic FCC, with the Cartesian axes oriented along the cube edges (Figure \ref{fig:cp_basis}), the non-zero elements of $\CC$ are the same ones as for the simplest anisotropic solid, the three values $c_{11}$, $c_{12}$ and $c_{44}$ are independent.  The elastic constants are chosen to be  $c_{11} = 280.5 \text{GPa}$, $c_{12} = 188.5 \text{MPa}$, and $c_{44} = 133.5 \text{GPa}$, see \eqref{C}. The reference values are as follows: the length $l_0 = 10\mu\, m$, the velocity $\vv_0 = 1\mu m/s$, the reference slip rate $c_0 = 0.1 s$, the critical resolved shear stress $\tau_c = 420 \text{MPa}$.  The rate sensitivity parameter $m$ is taken to be $0.05$, for $\beta = \frac{1}{m}$, compare \eqref{alpha slip}. A review of computational strategies in an effort to overcome numerical instabilities connected with the power law in (\ref{alpha slip}) can be found e.g. in \cite{DelJa:06,KuRad:08}. In accordance with the assumptions of our approach, the hardening effects are neglected. These material parameters are those for a single-crystal Ni-base superalloy, which has the nominal composition of NiָCoַCrַTaֶ.2Alֵ WֳReֲMoְ.2Hf, see~\cite{Shade2009}.

We solve the set of equations derived in Section \ref{sec:model}, supplemented with proper boundary and initial conditions, see also \eqref{finalsystem}. Let $T>0$, $\Omega\subset \mathbb{R}^{3}$ open and bounded, $\rho_0:\Omega \to \Re$, $\vv_0:\Omega \to \mathbb{R}^{3}$, $\S_0:\Omega \to \Re^{3\times 3}$, $\ve{a_i}_0:\Omega \to \mathbb{R}^{3},\quad i=1,2,3$, $\vv_D:\Gamma_D \to \mathbb{R}^{3}$, $\ve{t}:\Gamma_N \to \mathbb{R}^{3}$. The system is satisfied inside the domain $\Omega \subset \mathbb{R}^{3}$, the boundary $\partial \Omega$ consists of two non-intersecting parts $\Gamma_D$ and $\Gamma_N$ corresponding to the Dirichlet and the Neumann boundary conditions, respectively. We look for the density $\varrho$, the velocity $\vv$, the Kirchhoff stress $\S$, and the lattice vectors $\ve{a_1}$, $\ve{a_2}$, $\ve{a_3}$ ($\s$ and $\m$ are given through lattice vectors, equations \eqref{reciprocal} and \eqref{slipsystem}) satisfying

\begin{subequations}\label{compsystem}
\begin{align}
\dot \varrho +\varrho\, \text{div}\,\bs{v} &= 0,\label{compsystema} \\
\varrho\, \dot{\bs{v}}  &=  \text{div}\,(\varrho \S)\,, \label{compsystemb}\\
\ut{\S} &= \CC\left(\D -\sum_{\alpha=1}^{N}\nu^{(\alpha)}\sym\left(\s\otimes\m\right) \right) \,, \label{compsystemc} \\
\sum_{i=1}^3 \dot{\bs{a}}_{\ve{i}}
\otimes \bs{a}^{\ve{i}}  &=  \nabla \bs{v} - \sum_{\alpha=1}^{N}\nu^{(\alpha)}\left(\s\otimes\m\right)\,, \label{compsystemd} \\
\rho(x,0) &= \rho_0(x)  &\forall x \in \Omega,\\
\vv(x,0) &= \vv_0(x)  &\forall x \in \Omega,\\
\S(x,0) &= \S_0(x) &\forall x \in \Omega,\\
\ve{a_i}(x,0) &= \ve{a_i}_0(x)&\forall x \in \Omega,\\     
\vv &= \vv_D &\text{ on } \Gamma_D\times (0,T),\\
\S \ve{n} &= \ve{t} &\text{ on } \Gamma_N \times (0,T),
\end{align}
\end{subequations}
where $\rho_0(x) = 1$,  
$\S_0(x)\ve{n} = \ve{0} \text{ for } x \in\Gamma_N \times [0,\infty)$,
$\S_0(x) =  \ve{0} \text{ for } x \in \Omega$,
$\ve{a_{1}}_0(x) = (1,0,0)$, $\ve{a_{2}}_0(x) = (0,1,0)$, $\ve{a_{3}}_0(x) = (0,0,1)$ and $\ve{n}$ is the outer normal vector. 
Moreover, as already explained the slip rates $\nu^{(\alpha)}$ are governed by the equations (\ref{alpha slip}), (\ref{tau}).

Two kinds of top velocity boundary condition are considered, see Figure \ref{fig:3dcomp} (b-c). In the first case, corresponding to Kuroda's condition BC1 \cite{Kuroda2013}, we fix the velocity on the top boundary $\Gamma_1$ and require \begin{equation}
\vv_0(x,t)|_{\Gamma_1} = (0,0,-1).\label{bc1}
\end{equation}
This condition is referred below to as \emph{the constrained case} and the setting mimics the situation with a very stiff loading axis and the friction between the top surface and the plunger is extremely high. The second case (corresponding to Kuroda's condition BC2 \cite{Kuroda2013}) is called below \emph{the unconstrained case} and it allows the top surface to move in the plane perpendicular to the $z$-axis, which means that we require that 
\begin{equation}
\vv_0\cdot \ve{n} = \vv_{0_{z}}(x,t)|_{\Gamma_1} = -1 \quad \textrm{ and } \quad 
\S\ve{n} \cdot \ve{\tau_{i}} = 0, \quad i=1,2\,,  \label{bc2}
\end{equation}
where $\ve{\tau_1}$,  $\ve{\tau_2}$  are two vectors generating the basis of the tangent plane orthogonal to $\ve{n}$ at the boundary. This simply mimics the case where there is no friction between the top surface and the plunger. In the rest of the Dirichlet boundary, namely the bottom of the specimen, the velocity is fixed $\vv_0(x,t)|_{\Gamma_3} = (0,0,0)$.  The degrees of freedom for the velocity on the conical base are fully constrained. Moreover, we put the Neumann boundary condition $\S(x,t)\ve{n} = \ve{0}$ on the lateral surface $\Gamma_2$.

The crystal orientation is relative to the fixed sample coordinates in the reference configuration. 
The $[\bar 1 2 3]$ direction is set to be parallel to the sample axis, i.e. the $X_3$ direction, the $[1 \bar 1 1]$ direction is chosen to coincide with the $X_1$ direction, and the $X_2$ direction coincides with $[5 4 \bar 1]$ (the overbar denotes the negative value). The specification of the slip systems is given in Table \ref{tab:slipsystems}. The primary slip system is number 2, $(1 1 1)[ 1 0 \bar 1]$, the angle between $\ve{s}^{(2)}$ and $X_1-X_2$ plane reads $49.1^\circ$.

\begin{table}[ht]
\begin{center}
\begin{tabular}{|lcc|ccc|ccc|}
\hline 
$\alpha$ & $\sqrt{2}\s$  & $\sqrt{3}\m$ & $\alpha$ & $\sqrt{2}\s$  & $\sqrt{3}\m$  & $\alpha$ & $\sqrt{2}\s$  & $\sqrt{3}\m$ \\ 
\hline 
$1$ & $[\bar 1 1 0]$ & $(1 1 1)$ & $5$ & $[1 0 1]$ & $(1 \bar 1 \bar 1)$  & $9$ & $[0 \bar 1 \bar 1 ]$ & $(\bar 1 1 \bar 1)$\\ 
$2$ & $[ 1 0 \bar 1]$ & $(1 1 1)$ & $6$ & $[0  1 \bar 1]$ & $(1 \bar 1 \bar 1)$& $10$ & $[ 1 \bar 1 0]$ & $(\bar 1 \bar 1  1)$\\ 
$3$ & $[0 \bar 1 1 ]$ & $(1 1 1) $& $7$ & $[ 1  1 0]$ & $(\bar 1 1 \bar 1)$& $11$ & $[ \bar 1 0 \bar 1]$ & $(\bar 1 \bar 1  1)$\\  
$4$ & $[\bar 1 \bar 1 0]$ & $(1 \bar 1 \bar 1)$ & $8$ & $[ \bar1 0 1]$ & $(\bar 1 1 \bar 1)$ & $12$ & $[ 0 1 1]$ & $(\bar 1 \bar 1  1)$\\ 
\hline 
\end{tabular} 
\end{center}
\caption{The slip directions and normals for a FCC crystal in Miller notation.}
\label{tab:slipsystems}
\end{table}

\subsection{Finite element formulation}

In the solution of the micropillar boundary-value problem without higher-order gradients, Kuroda \cite{Kuroda2013} used a generally adopted finite element Lagrangian framework.
In his numerical experiments, he evaluated deformed finite element meshes, contour maps of slip, maps of lattice rotations, and stress-strain diagrams for nominal compressive strain up to $0.2$. In the present example, we employ  the Arbitrary Lagrangian Eulerian (ALE) approach in the sense that we use an Eulerian formulation on a moving mesh, which captures the free boundary of the micropillar. In this way, we are able to extend the tested range of  nominal compressive strain up to $0.5$. For a brief description of our numerical scheme we refer to the Appendix. The detailed description concerning accuracy and efficiency of the numerical method and further numerical results of the application of the ALE method to three-dimensional crystal plasticity problems will be described in detail in a forthcoming work.

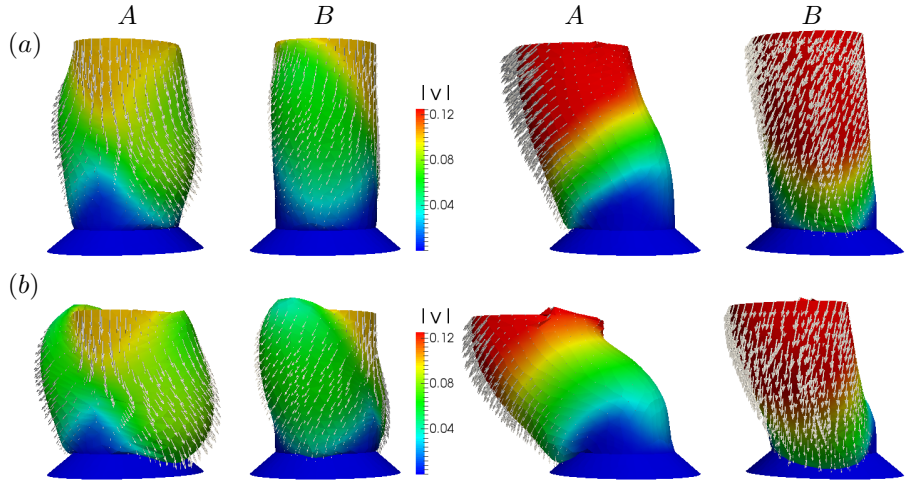
\begin{figure}[!h]
\begin{center}
\setlength{\unitlength}{.1\textwidth}
      \begin{picture}(10,5.25)
        \put(0,0){\includegraphics[width=\textwidth]{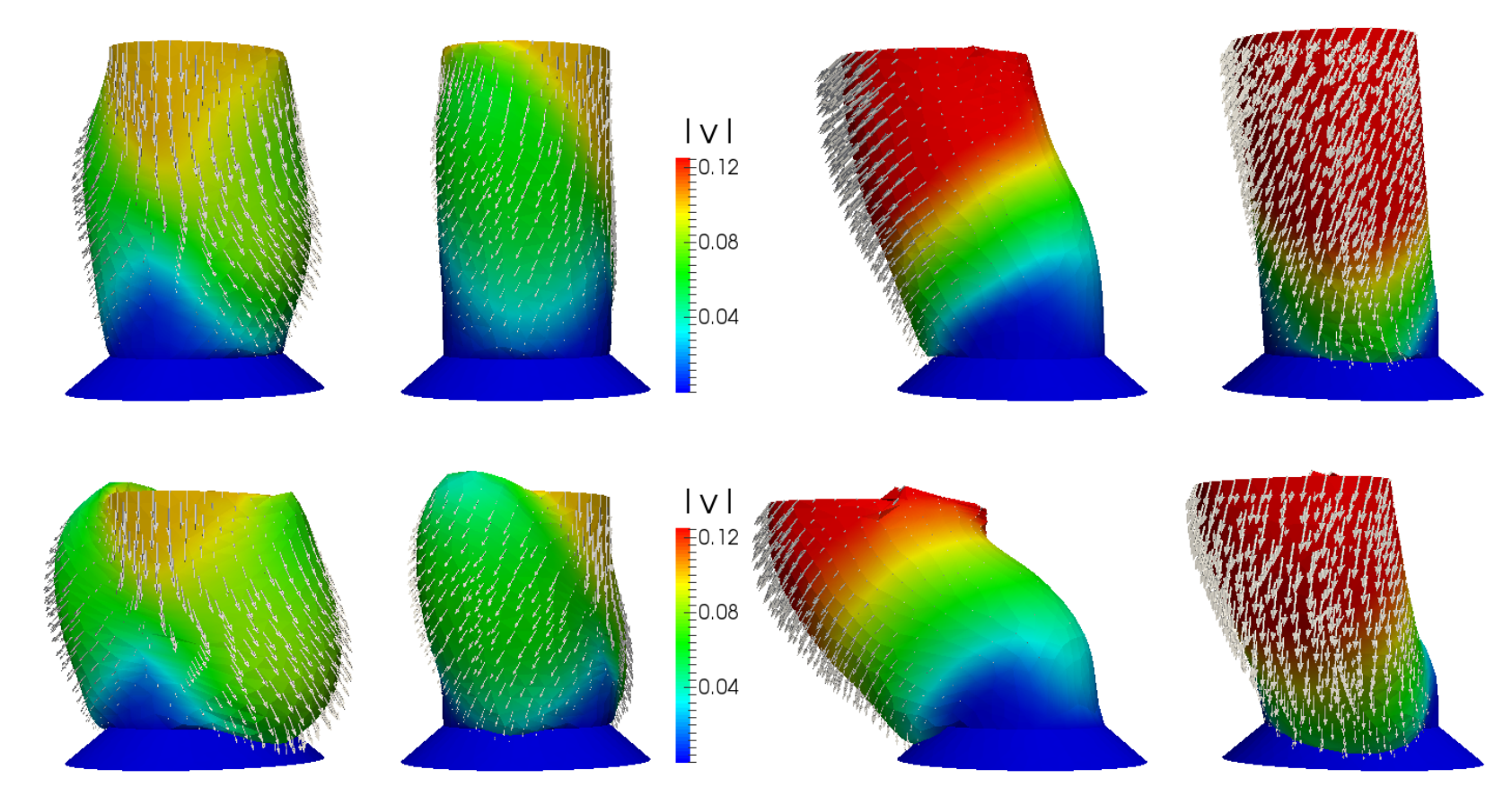}}
        \put(0,4.85){$(a)$}
        \put(0,2.2){$(b)$}
        \put(1.2,5.15){$A$}
        \put(3.35,5.15){$B$}
        \put(6.1,5.15){$A$}
        \put(8.7,5.15){$B$}
        \end{picture}
\caption{The velocity field (on the sample surface) for $X_2-X_3$ and $X_1-X_3$  plane views (denoted by A and B, respectively) with constrained (left) and unconstrained (right) boundary conditions for nominal strains 0.2 (a) and 0.4 (b).}
\label{fig:3dcomp_v}
\end{center}
\end{figure}

\subsection{Results}\label{sec:miccompres}

\begin{figure}[h!]
\setlength{\unitlength}{.1\textwidth}
\begin{center}
      \begin{picture}(10,8)
        \put(0.,0){\includegraphics[width=\textwidth]{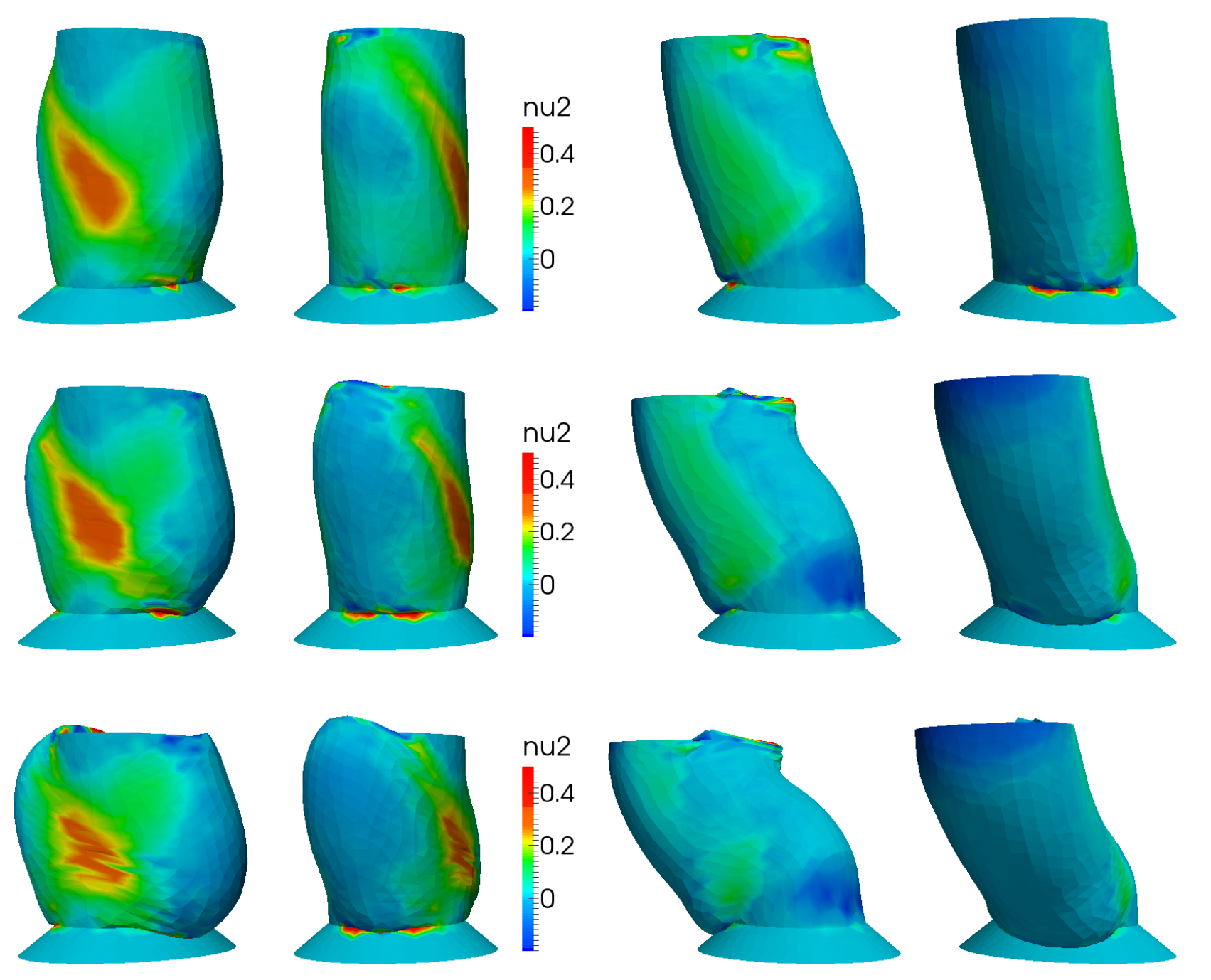}}
        \put(0,7.9){$(a)$}
        \put(0,5){$(b)$}
        \put(0,2.2){$(c)$}
        \put(1,8.1){$A$}
        \put(3.15,8.1){$B$}
        \put(6,8.1){$A$}
        \put(8.45,8.1){$B$}
     \end{picture}
\caption{Contours of slip, on the specimen surface, on the primary system ($\ve{s}^{(2)}$) for $X_2-X_3$ and $X_1-X_3$ plane views (denoted by A and B, respectively) with constrained (left) and unconstrained (right) boundary conditions for nominal strains 0.2 (a). 0.3 (b) and 0.4 (c).}
\label{fig:s2}
\end{center}
\end{figure}
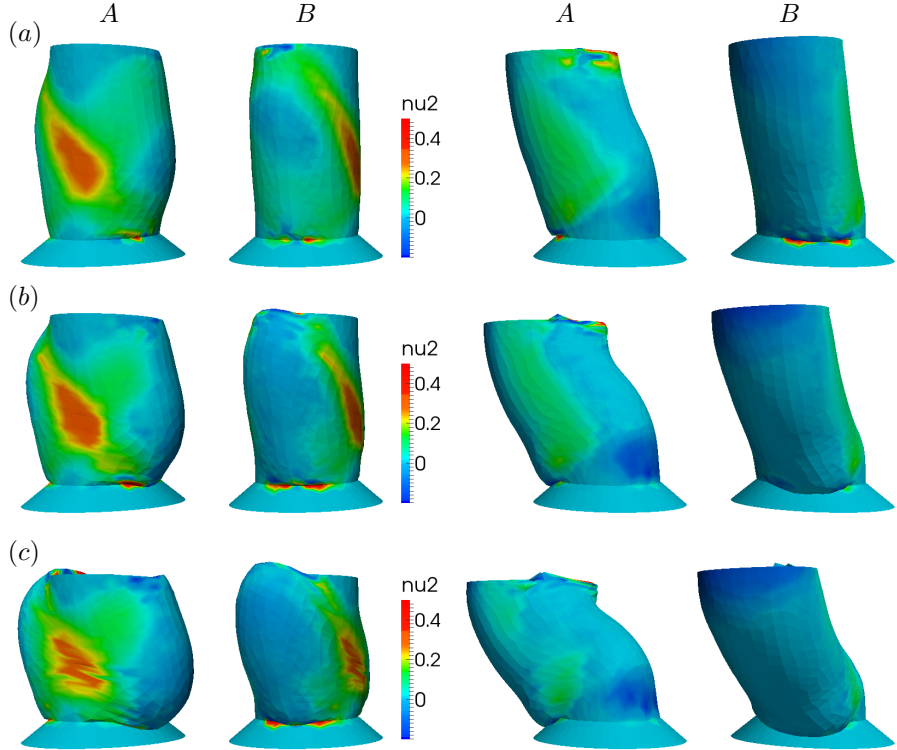

Since the computations are conducted in the Eulerian coordinates our primary variable is the velocity. In Figure \ref{fig:3dcomp_v} we present the velocity field at various strain levels in the whole sample for the $X_1-X_3$ and $X_2-X_3$ plane views. Due to anisotropy of the acting slip systems the deformation profiles are not the same for different plane views, Fig. 3 of \cite{Kuroda2013} exhibits the same features. We focus on the effect of the applied boundary conditions in comparison with a scanning electron microscope compression experiments, see \cite{Shade2009} and the numerical studies \cite{Kuroda2013,Shade2009}. The presented approach allows for nominal strains up to $0.5$ and overcomes numerical difficulties encountered in \cite{Kuroda2013,Shade2009}, where the authors report results for the nominal strain up to $0.2$. Nevertheless, for nominal strains higher than $0.5$ a collapse of our calculations appeared. The amount of mesh distortion that can be handled by the ALE method is limited and a full remeshing of the problem is eventually required to overcome this limit. % Note that the real material cracks ( before reaching strain of 0.5) what is beyond the scope of the model.

\subsection{Activity of slip systems}

In Figure \ref{fig:s2} we present the contour maps of slip on the primary system ($\ve{s}^{(2)}$) and compare the results for two kinds of boundary conditions. The sample deformation and the primary slip activity exhibit the same features as the one shown in Figure 3 of \cite{Kuroda2013}, in the nominal strain region up to 0.2. There is a measurable difference between the alignment of slip activity regions (shear bands) between different types of boundary conditions, see Figure~\ref{fig:s2}. The majority of deformation has been localized to an intense slip band. Slips are highly concentrated in a shear zone.

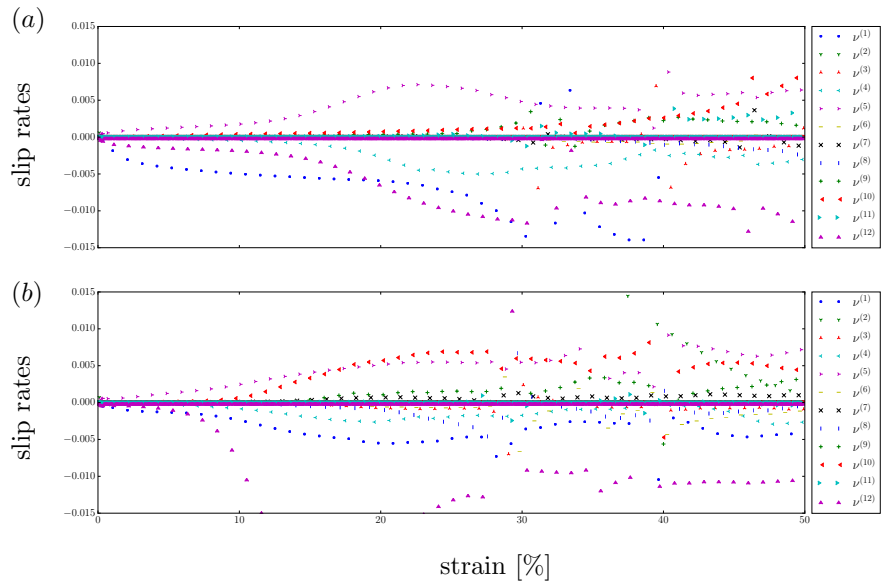
\begin{figure}[ht]
\setlength{\unitlength}{.1\textwidth}
\begin{center}
      \begin{picture}(10,6.25)
        \put(0,0){\includegraphics[width=\textwidth]{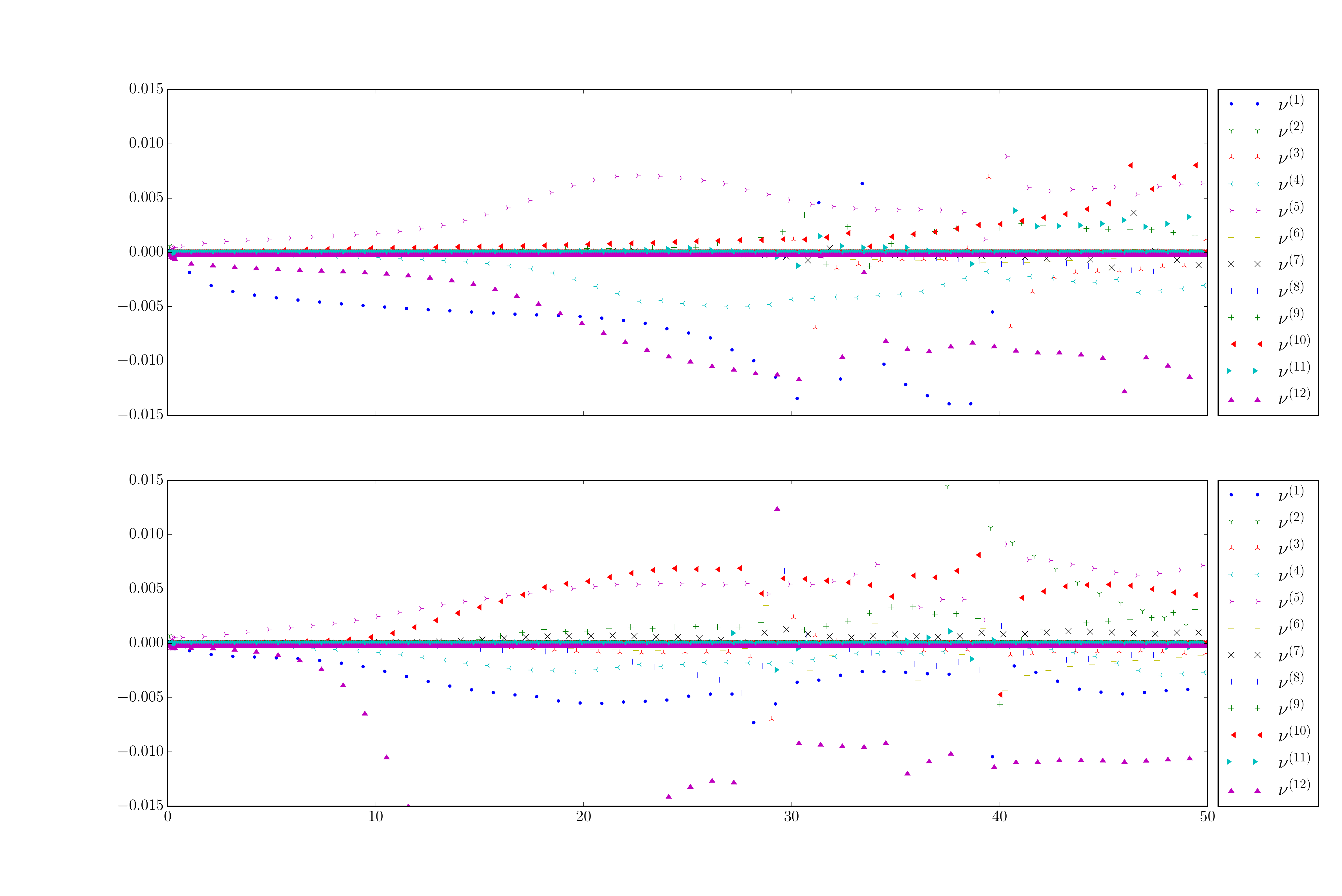}}
        \put(0.3,4.25){\rotatebox{90}{slip rates}}
        \put(0.3,6){$(a)$}
        \put(0.3,1.25){\rotatebox{90}{slip rates}}
        \put(0.3,3){$(b)$}
        \put(5,0){strain [\%]}
     \end{picture}
\caption{The average slip rate for individual slip systems for a constrained (a) and an unconstrained domain (b).}
\label{fig:slipactive}
\end{center}
\end{figure}

In Figure \ref{fig:slipactive} we present the average slip rate for each of twelve individual slip systems, and compare two kinds of boundary conditions. The slip rates are not qualitatively similar for different types of boundary conditions. While the constrained boundary conditions give rise to stable constant-in-time values, the unconstrained case changes the behaviour for strains over~$0.05$. In both cases we observe instabilities for high strains over 0.3, see Figure~\ref{fig:s2}. The~inactive slip systems $\alpha \in \{7,8,9\}$ are oriented in such a way that the slip direction $\s$ is perpendicular to the compression axis, namely $\m\cdot (-1,2,3)=0$, see Table~\ref{tab:slipsystems}.

\begin{figure}[h]
\setlength{\unitlength}{.1\textwidth}
\begin{center}
\begin{picture}(5,5)
        \put(0.,0.){\includegraphics[scale=.4]{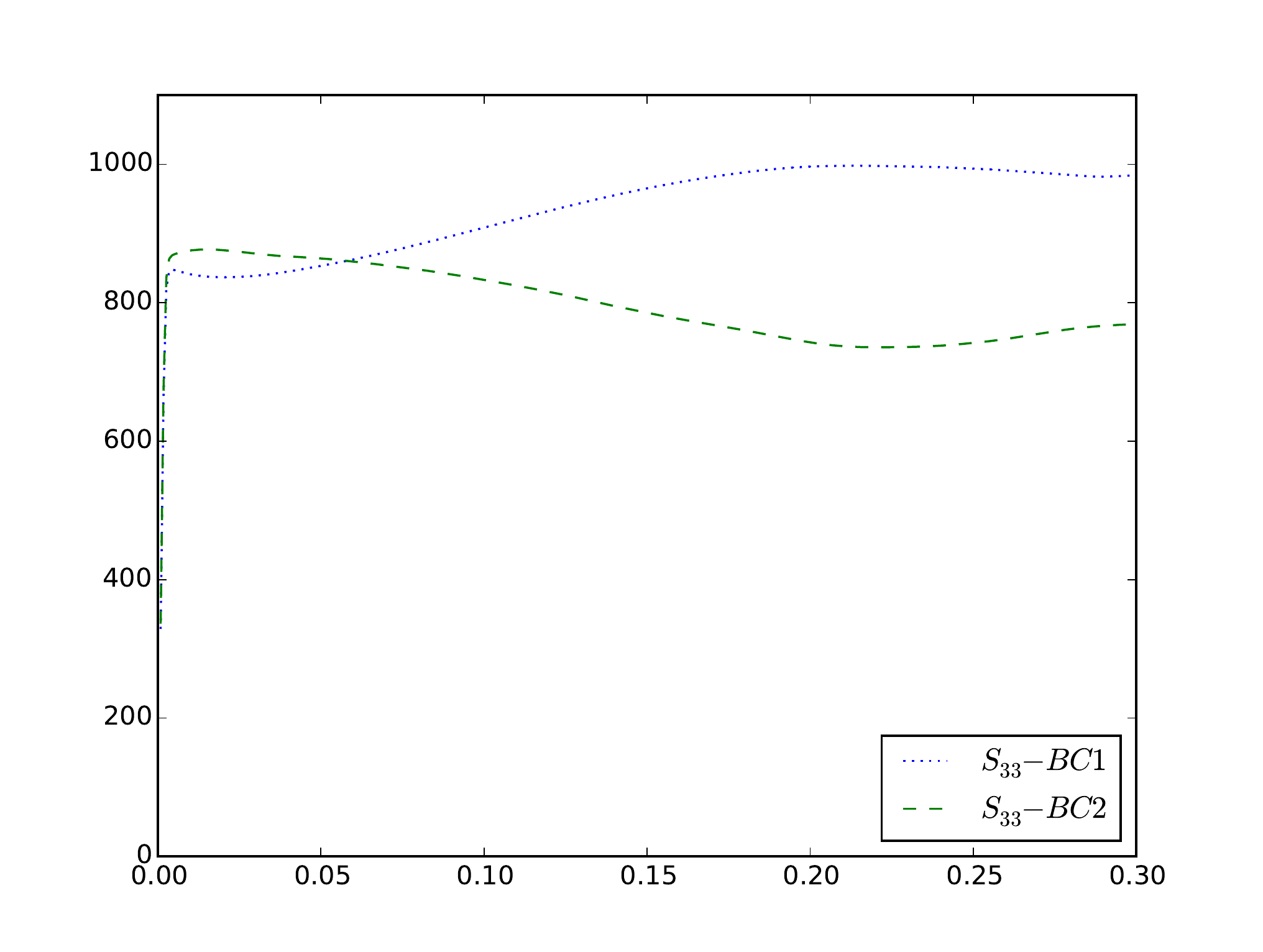}}
        \put(-0.4,0.4){\rotatebox{90}{Nominal compressive stress [MPa]}}
        \put(1.7,-.2){Nominal compressive strain [\%]}
\end{picture}
\end{center}
\caption{The nominal compressive stress for a constrained ($BC1$)  and an unconstrained domain ($BC2$).}
\label{fig:stressstrain}
\end{figure}

\subsection{Stress-strain behaviour}

Figure \ref{fig:stressstrain} shows the nominal compressive stress-nominal compressive strain curves.
Despite of the considered zero hardening, the constrained condition exhibited significant stress increase while the unconstrained condition showed a decrease of the calculated nominal compressive stress. The stress increase in the constrained case is caused by the Dirichlet velocity boundary condition on the top and bottom boundaries. There, plastic deformation is suppressed, and part of the sample deformed elastically causing the stress to increase. The effect is associated with barrelling
of the sample. On the other hand, the decrease of the nominal stress with the increasing nominal strain in the case of the unconstrained boundary condition is associated with the inclination of the sample.   
As is seen from the Figure \ref{fig:stressstrain} the results in the strain range up to $0.2$ agree qualitatively with the numerical results presented in \cite[Fig.11]{Kuroda2013} and \cite[Fig. 3(A)]{Shade2009}, and the experimental results \cite[Fig.4]{Shade2009}. However, to obtain an exact 
agreement with the experiments would require the application of boundary conditions that exactly mimic the experimental setting; this is however beyond of the scope of this work.

\section{Summary}\label{sec:con}

\begin{itemize} 

\item We presented a new approach to the derivation of a thermodynamically compatible
rate-type fluid model for crystal plastic materials. We employed a Gibbs-potential-based approach and derived a rate-type stress-strain constitutive equation.  Instead
of the standard crystal plasticity hypothesis that the plastic part of the
velocity gradient equals a sum of contributions of the individual slip
systems $ \Lp = \sum_{\alpha=1}^N \nu^{(\alpha)} \s \ot \m$ we have
employed the procedure of maximization of the rate of dissipation to
derive this relation from a single scalar function, \eqref{xi}. It is
worth emphasizing that the Gibbs-potential-based approach, as developed in \cite{RajSr:11} and presented here, automatically provides an objective rate derivative for the stress as well as the possibility to incorporate anisotropy of the material response.

\item We have incorporated an Eulerian form for the evolution of the lattice basis vectors instead of computing directly the evolution of the slip directions and the normals to the slip plane. This is one of the novel ingredients of the developed model. It allows us to perform efficient three-dimensional computations of the material with a full set of slip systems (e.g. 12 slip systems in the case of FCC structure), as illustrated on two test problems concerning three-dimensional micropillar compression.

\item We have formulated a finite element discretization scheme, which has been used to solve the fully coupled problem. Our solver is monolithic. The Arbitrary Lagrangian Eulerian (ALE) approach has been applied in order to use the Eulerian formulation on a moving mesh, that captures the free boundary. We developed a new algorithm for a three-dimensional compression and achieved qualitative correspondence between our results and the experimental and numerical results concerning the same problem, as presented in \cite{Kuroda2013}.

\item The fact that our method is based on the Eulerian formulation, and consequently the primal role is given to the velocity of the material and the distortion of the crystal lattice space, enabled us to demonstrate that the whole approach is capable of simulating deformation processes with large strains for suitably defined problems. 

\end{itemize}

\section*{Appendix}

The ALE approach is a finite element formulation in which the computational
system is not a priori fixed in space (Eulerian) or attached to the material (Lagrangian) \cite{Hirt1974,Hughes2007}. In fact we can formulate the problem on an arbitrary moving, time-dependent domain and define the ALE mapping between a fixed computational domain 
and an arbitrary domain. The ALE mapping is the key to transforming our system to a
fixed computational domain. To this end the proper definition of the ALE mapping is of a great importance. The class of ALE methods is widely applied, e.g. to fluid-structure interaction problems. In crystal plasticity, the ALE method has been successfully tested in the solution of a two-dimensional problem of a channel-die compression in \cite{CazIon:09,Cazacu2010,Cazacu2010a}, where the nominal compressive strain of $0.45$ has been reached.

In the present paper, the ALE method is used to solve the system \eqref{compsystem}. The implementation was done using FEniCS \cite{Blechta2015, fenics:book}. A mixed finite element discretization is used in space, and time is discretized by the backward Euler scheme. The choice of the conforming simplical tetrahedral elements for each variable consists of $\mathcal{P}_2$ elements for the velocity, $\mathcal{P}_1$ for the density and the lattice basis vectors, and $\mathcal{P}_1-disc$ (discontinuous elements) for the Kirchhoff stress. We employ the ALE approach in the sense that we use our Eulerian formulation on a moving mesh, which captures the free boundary.

To solve \eqref{compsystem} at each time step, we perform two actions. Given $\vv^{k}$, $\rho^{k}$, $\S^{k}$, $\ve{a_i}^{k}$ from the previous time level we proceed as follows:
\newline

\textbf{Step 1:} Solve problem for $\vv_h^{k+1}, \rho_h^{k+1}, \S_h^{k+1}, $ and $\ve{a_i}_h^{k+1}$ ($i \in \{1,2,3\}$)

\begin{align*}
\left(\frac{\rho_h^{k+1}-\rho_h^{k}}{\Delta t} + \div(\rho_h^{k+1} \vv_h^{k+1}) , z_\rho \right) &= 0,\\
  \rho_h^{k+1}\left(\frac{\vv_h^{k+1}	- \vv_h^k}{\Delta t} + \left(\vv_h^{k+1}-\vv^k_{h-\mathrm{mesh}}\right) \nabla \vv_h^{k+1},\ve{z_v} \right)  +  \left(\rho_h^{k+1} \S_h^{k+1} ,\nabla \ve{z_v} \right)  &= 0, \\
\left(\ut{\S}_h^{k+1} - \CC \De_h^{k+1} ,\te{Z_S} \right) &=  0, \\
\left( \frac{\ve{ a_i}_h^{k+1} - \ve{ a_i}_h^k}{\Delta t}  - \Le_h^{k+1} \ve{a_i}_h^{k+1}  ,\ve{z_{a_i}} \right) &=  0,
\end{align*}

where $$\Le_h^{k+1} = \nabla \vv_h^{k+1} - \sum_{\alpha=1}^{N}\nu_h^{(\alpha)k+1}\left(\ve{s}_h^{(\alpha)k+1}\otimes\ve{m}_h^{(\alpha)k+1}\right) \text{ and }\De_h = \sym \Le_h.$$

\textbf{Step 2:} Move the mesh by $\ve{u}_h^k = \vv^k_{h-\mathrm{mesh}} \Delta t$. The velocity of the mesh motion $\vv^k_{h-\mathrm{mesh}}$ is taken as the material velocity $\vv_h^k$, which in fact leads to an updated Lagrangian description.

The mesh velocity in the second step can be chosen arbitrarily, so that $\vv^k_{h-\mathrm{mesh}}|\partial \Omega = \vv^k_h|\partial \Omega$, which leads to the ALE description.

\section*{Acknowledgements}
The authors acknowledge the support of GACR [grant number P107/12/0121]. P. Minakowski also acknowledges the support of NSC (Poland) [grant number 2012/07/N/ST1/03369]. The authors thank Martin Kru\v{z}\'\i k for general discussions on this topic, Jaroslav Hron for specific discussions on numerics-related issues and Endre S\"{u}li for several suggestions improving the final form of the text. 
\section*{References}
\bibliography{reference}
\bibliographystyle{elsarticle-harv} 
\end{document}